\documentclass[twocolumn]{aa}
\usepackage{graphicx,epsfig}
\newcommand{\hi}{H\,{\sc i}}
\newcommand{\kms}{km\,s$^{-1}$}
\newcommand{\cmsq}{cm$^{-2}$}

\def\xp#1{{10$^{#1}$}}
\begin{document}
\title{Observations of \hi\ absorbing gas in compact radio sources at
cosmological redshifts}
\titlerunning{\hi\ absorption in distant radio sources}
\author{R.~C.~Vermeulen \inst{1}
        \and
        Y.~M.~Pihlstr\"om \inst{2,3}
        \and
        W.~Tschager \inst{4}
        \and
        W.~H.~de~Vries \inst{5,6,7}
        \and
        J.~E.~Conway \inst{3}
        \and
        P.~D.~Barthel \inst{5}
        \and
        S.~A.~Baum \inst{6}
        \and
        R.~Braun \inst{1}
        \and
        M.~N.~Bremer \inst{4,8}
        \and
        G.~K.~Miley \inst{4}
        \and
        C.~P.~O'Dea \inst{6}
        \and
        H.~J.~A.~R\"ottgering \inst{4}
        \and
        R.~T.~Schilizzi \inst{4,9}
        \and
        I.~A.~G.~Snellen \inst{10}
        \and
        G.~B.~Taylor \inst{2}
}
\offprints{R.C.~Vermeulen,\hfill\break {\tt rvermeulen@astron.nl}}
\institute{
         Netherl.\ Foundation for Research in Astronomy (ASTRON),
              P.O. Box 2, 7990 AA Dwingeloo, The Netherlands 
         \and 
             National Radio Astronomy Observatory, P.O. Box O,
             Socorro, NM 87801, U.S.A. 
         \and 
             Onsala Space Observatory, 439 92 Onsala, Sweden 
         \and 
             Leiden Observatory, PO Box 9513, 2300RA Leiden, The
             Netherlands 
         \and 
             Kapteyn Astronomical Institute, P.O. Box 800,
              9700 AV Groningen, The Netherlands 
         \and 
             Space Telescope Science Institute, Baltimore, U.S.A. 
         \and 
             Lawrence Livermore National Laboratories, U.S.A. 
         \and 
             Astrophysics Department, University of Bristol, Great Britain 
         \and 
             Joint Institute for VLBI in Europe (JIVE), Postbus 2, 7990
             AA Dwingeloo, The Netherlands 
         \and 
             Institute for Astronomy, University of Edinburgh,
             Blackford Hill, Edinburgh EH9 3HJ, United Kingdom 
}
\date{Received / Accepted }
\authorrunning{Vermeulen et al.}
\abstract{We present an overview of the occurrence and properties of
  atomic gas associated with compact radio sources at redshifts up to
  $z=0.85$. Searches for \hi\ 21\,cm absorption were made with the
  Westerbork Synthesis Radio Telescope at UHF-high frequencies
  (725--1200 MHz)\null. Detections were obtained for 19 of the 57
  sources with usable spectra (33\%). We have found a large range in
  line depths, from $\tau=0.16$ to $\tau\le0.001$. There is a
  substantial variety of line profiles, including Gaussians of less than
  10\,\kms, to more typically 150\,\kms, as well as irregular and
  multi-peaked absorption profiles, sometimes spanning several hundred
  \kms. Assuming uniform coverage of the entire radio source, we obtain
  column depths of atomic gas between $1\times 10^{19}$ and $3.3\times
  10^{21}$ ($T_{\rm sp}/100$ K)($1/f$) \cmsq. There is evidence for
  significant gas motions, but in contrast to earlier results at low
  redshift, there are many sources in which the \hi\ velocity is
  substantially negative (up to $v=-1420$\,\kms) with respect to the
  optical redshift, suggesting that in these sources the atomic gas,
  rather than falling into the centre, may be be flowing out,
  interacting with the jets, or rotating around the nucleus.

}
\maketitle

%
\section{Introduction \label{sec:intro}}

The energy output from AGN is commonly agreed to be powered by
accretion onto a super-massive black hole, resulting from the infall of
gas transported from the host galaxy down to the central regions (e.g.,
Rees \cite{rees84}). Recent detailed and statistical studies of Seyfert
galaxies (e.g., Cid Fernandes et al.\ \cite{cid01}) are beginning to
chart the evolutionary sequence that leads to the production of an
active galactic nuclei in late-type (Sd through S0) galaxies and the
effect of that evolution on the host galaxy (e.g., Storchi-Bergmann et
al.\ \cite{storchi01}). The picture that is emerging is one of a close
tie between the galaxy environment and galaxy-galaxy interactions
leading to a temporal sequence of circumnuclear starbursts and
accretion onto the nuclear black hole (e.g., Sanders et al.\
\cite{sanders88}).

However, while this picture is becoming well defined for Seyferts,
very little is known regarding the situation in powerful radio
galaxies. The mostly radio-quiet AGN in a large nearby galaxy sample
studied by Ho et al.\ (\cite{ho97}) often reside in early-type disk
galaxy hosts, many of which have spiral structures, and the amount of
gas needed to sustain the low-luminosity activity for the expected
lifetime is readily available. But looking at radio-loud AGN, Martel
et al.\ (\cite{martel99}) have found that at least 89\%\ of the hosts
of 46 low-redshift ($z<0.1$) 3CR radio galaxies are ellipticals. At
larger redshifts, Dunlop et al.\ (\cite{dunlop03}) have shown that the
host galaxies of radio-quiet and radio-loud quasars, as well as radio
galaxies, are massive ellipticals. These may of course themselves be
merger products of late type galaxies. True spiral hosts of powerful
radio galaxies seem to be very rare (see also Ledlow et al.\
\cite{ledlow98}).

The predominance of elliptical hosts is often thought to be puzzling,
since these are usually assumed to be gas poor. It should be noted
first that major galaxy merger products are likely places where gas
has lost angular momentum and can feed into the central engine. Apart
from that, Walsh et al.\ (\cite{walsh89}) have shown that in
ellipticals the amount of long-wavelength IR emission, indicative of
the presence of dust and gas, is correlated with the occurrence and
strength of a central radio source. More recently, it has been shown
that early-type galaxies can contain gas in both molecular form (see
e.g.\ Rupen \cite{rupen97}; Knapp \& Rupen \cite{knapp96}), and in
atomic form (see e.g.\ Bregman et al.\ \cite{bregman92}; Oosterloo et
al.\ \cite{oosterloo99}), and this gas could act as a fuel reservoir
for the central engine. Indeed, central structures of gas are
frequently observed in nearby radio sources; for instance, 17 of 19
nearby FRI radio galaxies observed with the HST by Verdoes Kleijn et
al.\ (\cite{verdoes99}) have dust structures in their centres. Such
dust disks are also reported in molecular (e.g\ in 4C12.50, Evans et
al.\ \cite{evans99}) and atomic lines, e.g.\ in NGC\,4261 (Van
Langevelde et al.\ \cite{vanlangevelde00}) and NGC\,1052 (Vermeulen et
al.\ \cite{rcv1052}). Furthermore, observations of nearby radio-loud
AGN by van Gorkom et al.\ (\cite{vangorkom89}) have directly shown, in
about 30\%\ of the galaxies studied, the presence of cold gas visible
through the \hi\ $\lambda=21$ cm line of neutral hydrogen in
absorption towards the central radio source. In a recent study,
Morganti et al.\ (\cite{morganti01}) have found evidence that in their
radio galaxy sample the \hi\ detections are related to circumnuclear
tori.

At cosmological redshifts, the galactic medium is largely
unexplored. But the recent upgrade of the Westerbork Synthesis Radio
Telescopes (WSRT) has opened up a new avenue of research. The array of
14 telescopes, with its new multi-channel correlator (DZB), offers
good sensitivity as well as a superior capability to reject external
interference compared to single dish telescopes. The wide band
UHF-high receivers, spanning 725--1200 MHz, enable studies of the
occurrence and kinematics of atomic gas seen by means of the
redshifted \hi\ line in absorption against radio sources almost out to
$z = 1$. We here present the combined results of a number of \hi\ line
searches performed at the WSRT. These projects had a variety of goals,
all related to the properties of cold gas in compact radio sources
with a total extent less than a few arcseconds, within the appropriate
redshift range, in the Northern sky, and with a flux density above a
few hundred mJy (in order to have a reasonable opacity detection
threshold). A majority of the sources in this paper are classified as
Gigahertz Peaked Spectrum (GPS) or Compact Steep Spectrum (CSS)
sources, believed to be intrinsically compact (e.g.\ Fanti et al.\
\cite{fanti95}, O'Dea \cite{odea98}). Further, a subset of the sources
in this paper are end-on classical doubles (including both quasars and
galaxies). As a result of coordination between the present authors, we
here present the complete observational results for the sources in the
original programmes for which usable spectra were obtained, plus some
statistics on the absorber depth and kinematics. In a follow-up paper,
using a more tightly defined sample of GPS and CSS sources including
data from the literature, Pihlstr\"om et al.\ (\cite{pihlstrom03})
discuss \hi\ absorption in the context of other properties, and show,
for example, the existence of an anti-correlation between absorption
depth and linear size of the radio source.


\section{WSRT observations and data analysis \label{sec:obs}}

This paper presents the combined results of a number of different WSRT
projects, performed between 1997 and 2001, using the UHF-high
receivers. While there were slight variations in the observational and
data reduction methods used, the description below adequately describes
the essentials. The data reduction was carried out using the NRAO {\sc
AIPS} and Caltech {\sc DIFMAP} packages, typically switching
back-and-forth to make use of their respective strong points.

Typically, each target was first observed for a few hours, in each of
two orthogonal linear polarisations, with 128 spectral channels
covering a 10 MHz wide observing band (the maximum available when the
surveys started), centred at the frequencies predicted for \hi\ based
on the optical redshifts (these have some uncertainty, as discussed in
Section~\ref{sec:det}). Doppler tracking was used; it ensures
that spectra taken separately at arbitrary times of the day and year
can be directly combined pixel-by-pixel. Most spectra were Hanning
smoothed online, chiefly to combat the spectral rippling effects of
strong narrow-band interference somewhere in the observing band. As a
result, the effective spectral resolution in the initial observations
is about 150 kHz.

The combined initial target list incorporated about 80 sources. Despite
the WSRT's impressive ability to suppress external interference, by
virtue of its being an interferometric array, and having a
multi-channel correlator, we have encountered a number of frequency
intervals in which the level of the interference was always too high to
obtain any useful data, even after trying to observe at different
times, with different telescope orientations (hour angles), or with
slightly shifted frequency bands. In the end, we pursued the
observations for 57 sources to the point where the spectra showed
either a line detection or gave a reasonable upper limit. This often
involved follow-up observations done with narrower, offset bandwidths
and/or twice the number of frequency channels (at the price of fewer
correlated baselines) in the pursuit of line candidates.

\begin{figure*}[htb!]
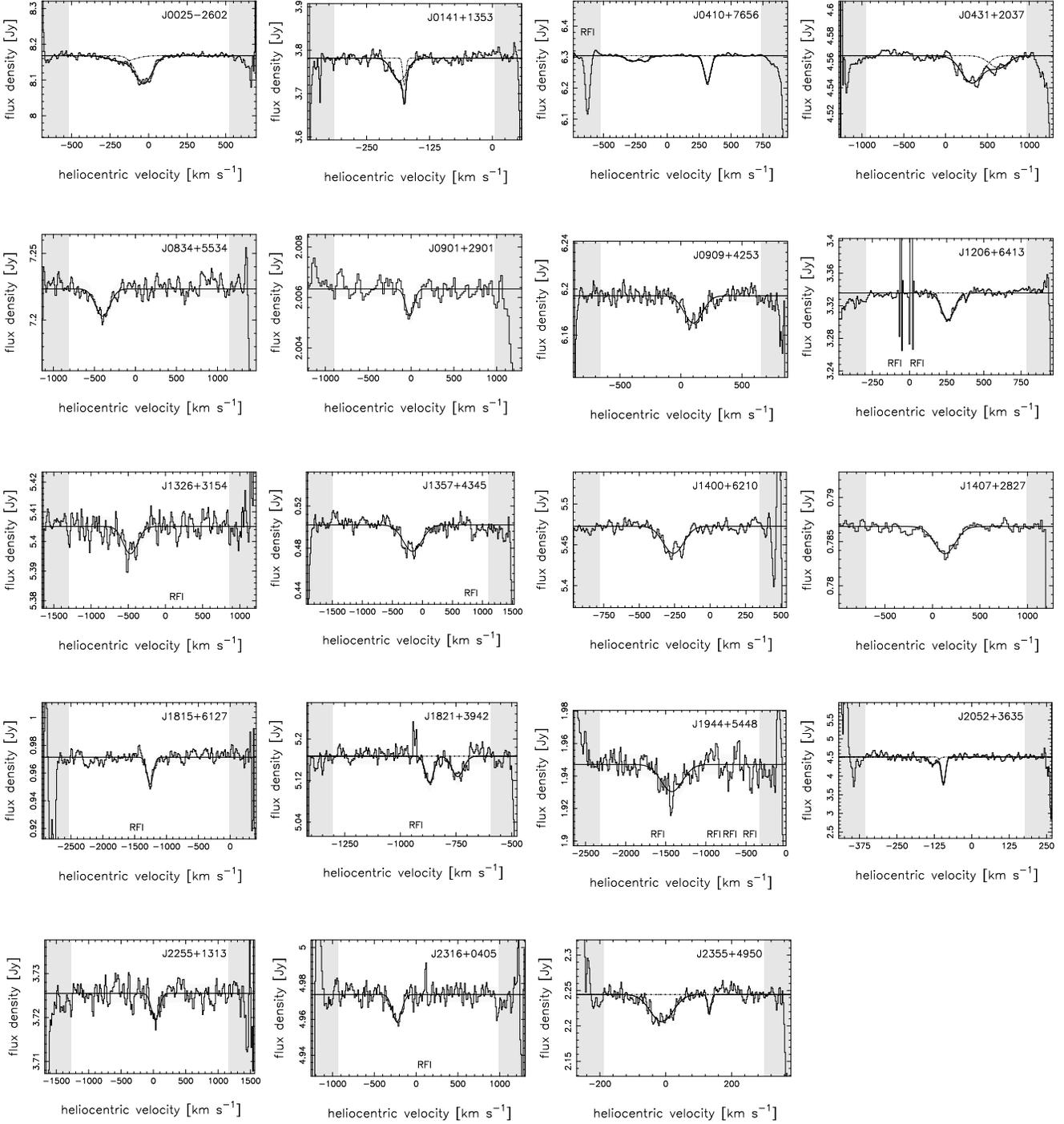

\centering
\centerline{
\includegraphics[angle=-90,width=4.2cm]{fig1a.ps}\hspace*{0.1cm}
\includegraphics[angle=-90,width=4.2cm]{fig1b.ps}\hspace*{0.1cm}
\includegraphics[angle=-90,width=4.2cm]{fig1c.ps}\hspace*{0.1cm}
\includegraphics[angle=-90,width=4.2cm]{fig1d.ps}\hspace*{0.1cm}
}
\vspace*{0.8cm}
\centerline{
\includegraphics[angle=-90,width=4.2cm]{fig1e.ps}\hspace*{0.1cm}
\includegraphics[angle=-90,width=4.2cm]{fig1f.ps}\hspace*{0.1cm}
\includegraphics[angle=-90,width=4.2cm]{fig1g.ps}\hspace*{0.1cm}
\includegraphics[angle=-90,width=4.2cm]{fig1h.ps}\hspace*{0.1cm}
}
\vspace*{0.8cm}
\centerline{
\includegraphics[angle=-90,width=4.2cm]{fig1i.ps}\hspace*{0.1cm}
\includegraphics[angle=-90,width=4.2cm]{fig1j.ps}\hspace*{0.1cm}
\includegraphics[angle=-90,width=4.2cm]{fig1k.ps}\hspace*{0.1cm}
\includegraphics[angle=-90,width=4.2cm]{fig1l.ps}\hspace*{0.1cm}
}
\vspace*{0.8cm}
\centerline{
\includegraphics[angle=-90,width=4.2cm]{fig1m.ps}\hspace*{0.1cm}
\includegraphics[angle=-90,width=4.2cm]{fig1n.ps}\hspace*{0.1cm}
\includegraphics[angle=-90,width=4.2cm]{fig1o.ps}\hspace*{0.1cm}
\includegraphics[angle=-90,width=4.2cm]{fig1p.ps}\hspace*{0.1cm}
}
\vspace*{0.8cm}
\hspace*{-4.4cm}
\includegraphics[angle=-90,width=4.2cm]{fig1q.ps}\hspace*{0.1cm}
\includegraphics[angle=-90,width=4.2cm]{fig1r.ps}\hspace*{0.1cm}
\includegraphics[angle=-90,width=4.2cm]{fig1s.ps}\hspace*{0.1cm}
\caption{Spectra of sources with detected \hi\ absorption. The fitted
Gaussian profiles are also shown, and features due to radio frequency
interference are marked 'RFI' on those plots. Velocities
are with respect to the optical redshifts. The grey horizontal bands
represent the 1$\sigma$ standard deviation from the fitted continuum
level. The grey vertical regions indicate the outer 25\% of the total
bandwidth.}
\label{fig:det}
\end{figure*}

\begin{figure*}[thb!]
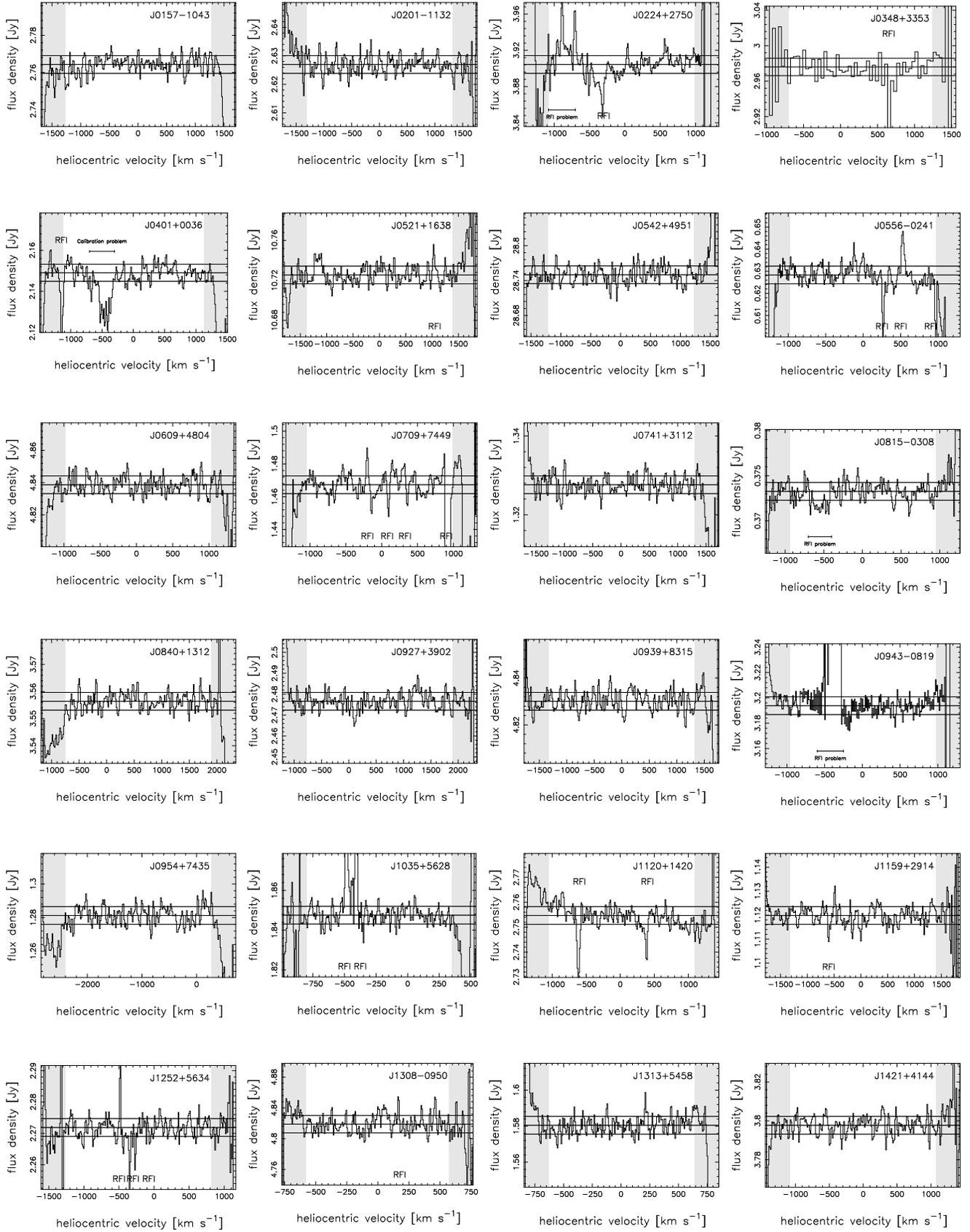

\centering
\centerline{
\includegraphics[angle=-90,width=4.2cm]{figa2a.ps}\hspace*{0.1cm}
\includegraphics[angle=-90,width=4.2cm]{figa2b.ps}\hspace*{0.1cm}
\includegraphics[angle=-90,width=4.2cm]{figa2c.ps}\hspace*{0.1cm}
\includegraphics[angle=-90,width=4.2cm]{figa2d.ps}\hspace*{0.1cm}
}
\vspace*{0.8cm}
\centerline{
\includegraphics[angle=-90,width=4.2cm]{figa2e.ps}\hspace*{0.1cm}
\includegraphics[angle=-90,width=4.2cm]{figa2f.ps}\hspace*{0.1cm}
\includegraphics[angle=-90,width=4.2cm]{figa2g.ps}\hspace*{0.1cm}
\includegraphics[angle=-90,width=4.2cm]{figa2h.ps}\hspace*{0.1cm}
}
\vspace*{0.8cm}
\centerline{
\includegraphics[angle=-90,width=4.2cm]{figa2i.ps}\hspace*{0.1cm}
\includegraphics[angle=-90,width=4.2cm]{figa2j.ps}\hspace*{0.1cm}
\includegraphics[angle=-90,width=4.2cm]{figa2k.ps}\hspace*{0.1cm}
\includegraphics[angle=-90,width=4.2cm]{figa2l.ps}\hspace*{0.1cm}
}
\vspace*{0.8cm}
\centerline{
\includegraphics[angle=-90,width=4.2cm]{figa2m.ps}\hspace*{0.1cm}
\includegraphics[angle=-90,width=4.2cm]{figa2n.ps}\hspace*{0.1cm}
\includegraphics[angle=-90,width=4.2cm]{figa2o.ps}\hspace*{0.1cm}
\includegraphics[angle=-90,width=4.2cm]{figa2p.ps}\hspace*{0.1cm}
}
\vspace*{0.8cm}
\centerline{
\includegraphics[angle=-90,width=4.2cm]{figa2q.ps}\hspace*{0.1cm}
\includegraphics[angle=-90,width=4.2cm]{figa2r.ps}\hspace*{0.1cm}
\includegraphics[angle=-90,width=4.2cm]{figa2s.ps}\hspace*{0.1cm}
\includegraphics[angle=-90,width=4.2cm]{figa2t.ps}\hspace*{0.1cm}
}
\vspace*{0.8cm}
\centerline{
\includegraphics[angle=-90,width=4.2cm]{figa2u.ps}\hspace*{0.1cm}
\includegraphics[angle=-90,width=4.2cm]{figa2v.ps}\hspace*{0.1cm}
\includegraphics[angle=-90,width=4.2cm]{figa2w.ps}\hspace*{0.1cm}
\includegraphics[angle=-90,width=4.2cm]{figa2x.ps}\hspace*{0.1cm}
}
\caption{Spectra of sources without detected \hi\ absorption. The
horizontal lines shown represent the $1\sigma$ standard deviation from
the fitted continuum level. Features due to radio frequency
interference are marked 'RFI'. Velocities are with respect to the
optical redshifts, positive corresponding to infall. The grey vertical
regions indicate the outer 25\% of the total bandwidth.}
\label{fig:nondet}
\end{figure*}

Bright external sources such as 3C48, 3C147, and 3C286 were used as
calibrators. Time-limited or baseline specific interference was first
removed, before using them for standard complex gain and
bandpass-calibration. No attempt was made to obtain polarisation
information; the data correlated in the two orthogonal linear
polarisations were simply added after separate gain calibration, to
increase the signal-to-noise of the final spectra. Furthermore, since
our primary interest was on line optical depths, no particular care was
expended on setting the overall flux density scale, but we believe that the
flux densities shown in Figures \ref{fig:det} and \ref{fig:nondet} are
accurate to better than 5\%.

\begin{table*}[htbp!]
\caption{Sources in which \hi\ absorption has been detected, and the
line properties. $z_{\rm opt}$ is the optical redshift, $\nu_{{\sf
HI}{\rm , det}}$ the centroid frequency of the detected \hi\
line, V$_{{\sf HI}{\rm , det}}$ the velocity offset of the detected line
centroid from the optical redshift (negative means blueshifted
line), $\tau_{\rm peak}$ the peak line optical depth, $\Delta V$ the
line FWHM, and finally $N_{\sf HI}$ the derived column depth.}
\label{tab:det}
\begin{center}
\begin{tabular}{lllcllrrrrr}
\hline
\hline\noalign{\medskip}
Source & Name & Other &  Optical & Radio\footnotemark[1] & 
$z_{\rm opt}$\footnotemark[2] & $\nu_{{\sf HI}{\rm , det}}$ &
V$_{{\sf HI}{\rm , det}}$ & $\tau_{\rm peak}$ & $\Delta V$ &
$ N_{\sf HI}$\footnotemark[3]\\
J2000 & B1950 & Name & ID & ID &   & MHz & \kms &
$10^{-2}$ &\kms&\xp{20}\,\cmsq\\\noalign{\medskip}
\hline\noalign{\medskip}
J0025$-$2602&0023$-$263&OB$-$238 &G&CSS & 0.322 & 1074.6 &  $-$30 & 0.93 & 126 & 2.14 \\ 
           &          &        &   &    &       & 1075.1 & $-$174 & 0.20 &  39 & 0.14 \\
J0141+1353 & 0138+136 & 3C49   & G &CSS & 0.621 &  876.7 & $-$175 & 1.66 &   7 & 0.22 \\
           &          &        &   &    &       &  876.7 & $-$185 & 1.39 &  35 & 0.88 \\
J0410+7656 & 0403+768 & 4C76.03& G &GPS & 0.5985&  887.6 &    315 & 1.40 &  61 & 1.55 \\
           &          &        &   &    &       &  889.4 & $-$275 & 0.30 & 107 & 0.58 \\
           &          &        &   &    &       &  889.1 & $-$170 & 0.27 &  77 & 0.38 \\
J0431+2037 & 0428+205 & OF247  & G &GPS & 0.219 & 1164.1 &    318 & 0.46 & 297 & 2.52 \\
           &          &        &   &    &       & 1162.9 &    636 & 0.21 & 247 & 0.93 \\
J0834+5534 & 0831+557 & 4C55.16& G &RG  & 0.242 & 1145.1 & $-$399 & 0.28 & 207 & 1.07 \\
J0901+2901 & 0858+292 & 3C213.1& G &CSS & 0.194 & 1189.7 &  $-$14 & 0.05 & 115 & 0.11 \\
J0909+4253\footnotemark[4]&0906+430&3C216&Q&CSS&0.670&850.2&    102 &0.38 & 177 & 1.23 \\ 
J1206+6413 & 1203+645 & 3C268.3& G &CSS &0.371  & 1035.1 &    258 & 1.00 & 101 & 1.85 \\
           &          &        &   &    &       & 1034.7 &    381 & 0.30 &  19 & 0.10 \\
J1326+3154 & 1323+321 & 4C32.44& G &GPS &0.370  & 1038.3 & $-$471 & 0.17 & 229 & 0.71 \\
J1357+4354 & 1355+441 &        & G &GPS &0.646\footnotemark[5]  &  863.4 & $-$165 & 5.00 & 367 &33.40 \\
J1400+6210 & 1358+624 & 4C62.22& G &GPS &0.4310 &  993.5 & $-$258 & 0.61 & 170 & 1.88 \\
J1407+2827 & 1404+286 & OQ208  & G &GPS &0.07658& 1318.9 &    131 & 0.39 & 256 & 1.83 \\
J1815+6127 & 1815+614 &        & Q &GPS &0.601  &  890.9 &$-$1258 & 2.03 & 118 & 4.35 \\
J1821+3942 & 1819+396 & 4C39.56& G &CSS &0.798  &  792.3 & $-$869 & 1.00 &  44 & 0.80 \\
           &          &        &   &    &       &  792.0 & $-$742 & 0.76 &  61 & 0.85 \\
J1944+5448 & 1943+546 & OV573  & G &GPS &0.263  & 1130.0 &$-$1420 & 0.86 & 315 & 4.91 \\
J2052+3635 & 2050+364 &        & G &GPS &0.355  & 1048.6 &  $-$95 &16.11 &  16 & 4.69 \\
           &          &        &   &    &       & 1048.7 & $-$130 & 4.40 &  32 & 2.56 \\
J2255+1313 & 2252+129 & 3C455  & Q &CSS &0.543  &  920.4 &     30 & 0.16 & 140 & 0.42 \\
J2316+0405 & 2314+038 & 3C459  & G &BLRG&0.2199 & 1165.3 & $-$229 & 0.31 & 130 & 0.72 \\
J2355+4950 & 2352+495 & OZ488  & G &GPS &0.2379 & 1147.5 &  $-$12 & 1.72 &  82 & 2.56 \\
           &          &        &   &    &       & 1147.2 &    133 & 1.16 &  13 & 0.28 \\
\noalign{\medskip}
\hline\noalign{\medskip}
\end{tabular}
\begin{tabular}{l}
$^1$ GPS = Gigahertz Peaked Spectrum source,
CSS = Compact Steep Spectrum source, RG = Radio Galaxy,\\ 
\hspace*{0.3cm}BLRG = Broad Line Radio Galaxy.\\
$^2$ The radio band centre frequency used during the observations
sometimes differed from that predicted by the optical\\
\hspace*{0.3cm}redshift, for example in order to avoid
RFI or to pursue tentative lines.\\
$^3$ Derived from $N_{\sf HI}=1.82\times10^{18} T_{\rm sp}
\tau_{\rm peak} \Delta V$ \cmsq, using $T_{\rm sp}=100$\,K.\\
$^4$ The detection of \hi\ absorption in 3C\,216, an integral part of
our survey, was first published in Pihlstr\"om et al.\ (\cite{pihlstrom99}.)\\
$^5$ Redshift based on Ca H$+$K and G-band stellar absorption features
in an unpublished spectrum obtained by some of us\\
\hspace*{0.3cm} (RCV and GBT) at the Palomar Observatory 200" Hale Telescope.\\
\end{tabular}
\end{center}
\end{table*}

After external calibration, interference was removed from the target
spectra to the extent possible. Figures \ref{fig:det} and
\ref{fig:nondet} show that this was not always completely successful:
some interfering signals, particularly when their strength did not
dominate in the observing band, could not be adequately discriminated
by their temporal or spatial characteristics. After editing to the
extent possible, those spectral channels without any line features were
averaged, and this continuum dataset was then used in iterative cycles
of self-calibration alternated with image cleaning and/or model-fitting
to the uv-data. The compact targets are all unresolved to the
WSRT. Characterisation of the continuum sources in the field was not a
goal in itself, and was pursued only to the level needed to find
adequate self-calibration gain factors, and to ensure that confusing
components would not disturb our spectral results. The self-calibration
complex gains found for the continuum were subsequently applied to all
spectral line channels. Final spectra were then produced by coherently
integrating all phase-calibrated visibilities per spectral channel;
this method produces an optimal signal-to-noise ratio given that in all
cases the target, at the phase centre, is by far the dominant source in
the field.

We find \hi\ absorption to be present in the spectra of 19 of the 57
sources (33\%); these are shown in Fig.~\ref{fig:det}. Table~\ref{tab:det}
gives the particulars of the detected absorption lines, determined by
Gaussian profile fitting of the peak optical depths, $\tau_{\rm peak}$,
and FWHM line widths, $\Delta V$; the column densities were computed
from these assuming uniform coverage of the entire radio source and a
spin temperature $T_{\rm sp}=100$\,K, using $N_{\rm
HI}=1.82\times10^{18} T_{\rm sp} \tau _{\rm peak} \Delta V$. Both
assumptions mean that the listed column depths are lower limits: the
covering factor might be below unity, and if the \hi\ absorption were
to arise in the pc-scale vicinity of the AGN, conditions might be such
that $T_{\rm sp} \simeq 8000$\,K (Maloney et al.\ \cite{maloney96}). For the
other 38 sources, we show the spectra in Fig.~\ref{fig:nondet}. For
Table~\ref{tab:nondet}, we have used a FWHM of 100\,\kms\ as a
plausible width for an absorption line in these objects (see
Section~\ref{subsec:prop-linew} and Fig.~\ref{fig:fwhm}), in order to
be able to derive $2\sigma$ upper limits to the line depths and \hi\
column depths, based on the continuum flux densities and r.m.s.\ noise
levels of the spectra.


\section{Detection uncertainties \label{sec:det}}

We expect that many of the optical redshifts used to centre the initial
observing bands are uncertain at the level of a few hundred \kms;
particularly at the higher redshifts, they are typically based on
emission lines which can be quite broad, and may have a velocity
centroid which is offset from the systemic velocity of the host galaxy.
With a 10\,MHz wide band, we have spanned a velocity range of
approximately $\pm 1250$\,\kms\ at $z=0.2$, increasing to approximately
$\pm 2000$\,\kms\ at $z=0.85$. Also note that an uncertainty of $\pm
0.001$ in $z$, which may be typical for the optical redshifts in the
literature, corresponds to a frequency uncertainty for \hi\ of nearly
$\pm1$\,MHz at $z=0.2$ but only about $\pm 0.4$\,MHz at $z=0.85$. Thus,
(optical) redshift uncertainties give the greatest (radio) frequency
uncertainties at the lowest redshifts. In view also of interpreting the
offsets discussed in Section~\ref{subsec:prop-flow}, it would be
valuable to obtain accurate systemic redshifts from high sensitivity,
high resolution optical spectroscopy, for all of the sources we have
detected in \hi. Nevertheless, Fig.~\ref{fig:vcen}, discussed in
Section~\ref{subsec:prop-flow}, shows that in the 19 sources with
detections, the \hi\ velocity distribution around the optical redshift
tails off rapidly towards $\pm500$\,\kms. We therefore believe that
our spanned bandwidth was adequate to cover the vast majority of the
potential lines, and quite likely all of them.

In view of the large range in flux densities amongst the targets, it
was impractical to achieve a uniform sensitivity to line optical depth
(opacity). It is difficult to quantify what this means for an
individual non-detection, but the following limited statistical
analysis is indicative. Figure~\ref{fig:sens} shows the individual
$2\sigma$ upper limits plotted along with the line opacities for the
detected lines, while Fig.~\ref{fig:cum} shows their cumulative
distributions. Firstly, we point out that from Fig.~\ref{fig:sens}
there appears to be no evidence for a strong correlation of optical
depth with FWHM, which means that adopting a somewhat arbitrary width
of 100\,\kms\ in order to compute upper limits is reasonable (the
limits would have been somewhat tighter had we adopted a larger
width). It is also clear that the distribution of detected line
strengths is different from that of the upper limits: many of the
limits are below most of the actual detections, meaning that these
limits are tight enough to be very significant. A KS test shows that
the cumulative distribution of the detections is different from that
of the 2$\sigma$ limits at the 98\% confidence level. It is also clear
from Fig.~\ref{fig:sens}, however, that about half of the 2$\sigma$
limits leave something to be desired: while they rule out the presence
of prominent absorption (roughly: above $\tau=1$\%), they still allow
the presence of shallower lines analoguous to those seen in some of
the weaker detections (roughly: below $\tau=0.5$\%). A significant
investment in telescope time would be needed in most cases in order to
constrain this potential incompleteness.

Nevertheless, the detections that we do have are already an important
next step in studying the cold neutral medium in cosmologically
distant radio sources. Below, we briefly discuss some of the line
properties. In a follow-up paper (Pihlstr\"om et al.\
\cite{pihlstrom03}), the line properties are related to other source
parameters. VLBI observations in order to image the neutral gas
distribution are already being pursued for some of the most prominent
absorbers (e.g., Vermeulen \cite{rcviau}, Vermeulen et
al. \cite{rcv2050}).

\section{Line Properties \label{sec:prop}}

\subsection{Line widths \label{subsec:prop-linew}}

Figure~\ref{fig:fwhm} shows the distribution of the FWHM of the
Gaussians fitted to the absorption lines. This plot confirms that \hi\
lines in these sources often are broad, with a FWHM of 100--200\,\kms\
or even more. We find that the primary lines (i.e.\ the deepest
Gaussians, with the highest optical depth in each source) are probably
somewhat broader than the secondary lines; the mean FWHMs, of
156\,\kms and 72\,\kms, respectively, differ at a confidence level of
97\%, according to a Student's $t$-test.

Often, \hi\ absorption is interpreted as being due to a circumnuclear
disk; for example VLA 21 cm \hi\ absorption imaging of a sample of
Seyfert and starbursts displayed almost exclusively sub kpc scale
rotating disks aligned with the outer galaxy disk (Gallimore et al.\
\cite{gallimore99}). Data for radio loud sources are also consistent
with disks on sub kpc scales, for example in Cygnus A (Conway
\cite{conway99}) and 1946+708 (Peck \& Taylor \cite{peck01}). Those
examples all have line widths of the order of 100\,\kms, comparable to
the detections reported in this paper. If the wide lines arise on sub
kpc scales close to the nucleus, the narrower lines could be
additional absorption occurring on larger scales within the galaxy,
provided that the line of sight does not cross a substantial amount of
differential rotation. Alternatively, the narrower lines could be from
single cohesive clouds, with only moderate internal velocity
dispersion, as is observed in 3C236 (Conway \&\ Schilizzi \cite{conway00}).

\begin{table*}[thp!]
\caption{Sources in which \hi\ absorption has not been detected, with
upper limits to the lines. $z_{\rm opt}$ is the optical redshift,
$\tau_{2\sigma}$
the limit to the optical depth for a line of width $\Delta V =
100$\,\kms\ centred at the optical redshift, and finally $N_{\sf
HI,2\sigma}$ the corresponding limit on the column depth. }
\label{tab:nondet}
\begin{center}
\begin{tabular}{lllcllrr}
\hline
\hline\noalign{\medskip}
Source & Name & Other & Optical & Radio\footnotemark[1]& 
$z_{\rm opt}$\footnotemark[2] & $\tau_{2\sigma}$ & 
$ N_{\sf HI,2\sigma}$\footnotemark[3]\\
J2000 & B1950 & name & ID & ID & & $10^{-2}$ & \xp{20} \cmsq\ \\ \noalign{\medskip}

\hline\noalign{\medskip}
J0157$-$1043 & 0155$-$109 & OC$-$192 & Q &EORQ& 0.616  & $<$0.35 & $<$0.63 \\
J0201$-$1132 & 0159$-$117 & 3C57     & Q &EORQ& 0.669  & $<$0.24 & $<$0.43 \\
J0224+2750   & 0221+276   & 3C67     & G &CSS & 0.3102 & $<$0.50 & $<$0.91 \\
J0348+3353   & 0345+337   & 3C93.1   & G &CSS & 0.243  & $<$0.48 & $<$0.87 \\
J0401+0036   & 0358+004   & 3C99     & G &EORG& 0.426  & $<$0.39 & $<$0.71 \\
J0521+1638   & 0518+165   & 3C138    & Q &CSS & 0.759  & $<$0.17 & $<$0.30 \\
J0542+4951   & 0538+498   & 3C147    & Q &CSS & 0.545  & $<$0.11 & $<$0.19 \\
J0556$-$0241 & 0554$-$026 &          & G &GPS & 0.235  & $<$2.17 & $<$3.94 \\
J0609+4804   & 0605+480   & 3C153    & G &EORG& 0.2769 & $<$0.23 & $<$0.41 \\
J0709+7449   & 0702+749   & 3C173.1  & G&FRII & 0.2921 & $<$0.74 & $<$1.34 \\ 
J0741+3112   & 0738+313   & OI363    & Q &GPS & 0.635  & $<$0.34 & $<$0.62 \\
J0815$-$0308 & 0812$-$029 & 3C196.1  & G &EORG& 0.198  & $<$0.53 & $<$0.96 \\
J0840+1312   & 0838+133   & 3C207    & Q &RQ  & 0.6808 & $<$0.18 & $<$0.33 \\
J0927+3902   & 0923+392   & 4C39.25  & Q &EORG& 0.6948 & $<$0.34 & $<$0.62 \\
J0939+8315   & 0931+834   & 3C220.3  & G &EORG& 0.685  & $<$0.17 & $<$0.31 \\
J0943$-$0819 & 0941$-$080 &          & G &GPS & 0.228  & $<$0.44 & $<$0.80 \\
J0954+7435   & 0950+748   &          & G &RG  & 0.695\footnotemark[4]  & $<$0.87 & $<$1.58 \\ 
J1035+5628   & 1031+567   & OL553    & G &GPS & 0.459  & $<$0.48 & $<$0.87 \\
J1120+1420   & 1117+146   & 4C14.41  & G &GPS & 0.362  & $<$0.21 & $<$0.38 \\
J1159+2914   & 1156+295   & 4C29.45  & Q &EORG& 0.729  & $<$0.66 & $<$1.20 \\
J1252+5634   & 1250+568   & 3C277.1  & Q &CSS & 0.321  & $<$0.25 & $<$0.45 \\
J1308$-$0950 & 1306$-$095 & OP$-$010 & G &CSS & 0.464  & $<$0.47 & $<$0.85 \\
J1313+5458   & 1311+552   &          & Q &RQ  & 0.613  & $<$0.62 & $<$1.13 \\
J1421+4144   & 1419+419   & 3C299    & G &CSS & 0.367  & $<$0.24 & $<$0.44 \\
J1443+7707   & 1443+773   & 3C303.1  & G &CSS & 0.267  & $<$0.52 & $<$0.94 \\
J1540+1447   & 1538+149   & 4C14.60  & Q &EORG& 0.605  & $<$0.22 & $<$0.39 \\
J1546+0026   & 1543+005   &          & G &GPS & 0.550  & $<$0.36 & $<$0.66 \\
J1642+6856   & 1642+690   & 4C69.21  & Q &EORG& 0.751  & $<$0.46 & $<$0.84 \\
J1658+0741   & 1655+077   & OS092    & Q &EORG& 0.621  & $<$0.54 & $<$0.98 \\
J1823+7938   & 1826+796   &          & G &GPS & 0.224  &$<$10.05&$<$18.29 \\ 
J1829+4844   & 1828+487   & 3C380    & Q &CSS & 0.692  & $<$0.07 & $<$0.12 \\
J1831+2907   & 1829+290   & 4C29.56  & G &CSS & 0.842  & $<$0.59 & $<$1.07 \\
J1845+3541   & 1843+356   & OU373    & G &GPS & 0.764  &$<$3.80 & $<$6.92 \\ 
J2022+6136   & 2021+614   & OW637    & Q &GPS & 0.227  & $<$0.13 & $<$0.24 \\ 
J2137$-$2042 & 2135$-$209 & OX$-$258 & G &CSS & 0.635  & $<$0.41 & $<$0.74 \\
J2250+1419   & 2247+140   & 4C14.82  & Q &CSS & 0.237  & $<$0.45 & $<$0.82 \\
J2321+2346   & 2318+235   & 3C460    & G &EORG& 0.268  & $<$0.70 & $<$1.28 \\
J2344+8226   & 2342+821   &          & Q &GPS & 0.735  & $<$0.26 & $<$0.47 \\
\noalign{\medskip}
\hline
\noalign{\medskip}
\end{tabular}
\begin{tabular}{l}
$^1$ GPS = Gigahertz Peaked Spectrum source, CSS = Compact Steep
Spectrum source, RG = Radio Galaxy,\\ \hspace*{0.3cm}RQ = Radio
Quasar, BLRG = Broad Line Radio Galaxy, FRII = Fanaroff-Riley Type 2,\\ 
\hspace*{0.3cm}EORG = End On Radio Galaxy and EORQ = End On Radio Quasar.\\
$^2$ The radio band centre frequency used during the observations
sometimes differed from that predicted by the optical\\
\hspace*{0.3cm}redshift, for example in order to avoid
RFI or to pursue tentative lines.\\
$^3$ Derived from $ N_{\sf HI,2\sigma}=1.82\times10^{18} T_{\rm sp} \tau_{2\sigma}
\Delta V$ \cmsq, using $T_{\rm sp}=100$\,K and $\Delta V = 100$\,\kms .\\
$^4$ Redshift based on Mg{\sc II}, [O{\sc II}], and [O{\sc III}] emission
lines in an unpublished spectrum obtained by some of us (RCV\\
\hspace*{0.3cm}and GBT) at the Palomar Observatory 200" Hale Telescope.\\
\vspace*{1.5cm}
\end{tabular}
\end{center}
\end{table*}

\addtocounter{figure}{-1}
\begin{figure*}[htb!]
\centering
\vspace*{0.8cm}
\centerline{
\includegraphics[angle=-90,width=4.2cm]{figb2a.ps}\hspace*{0.1cm}
\includegraphics[angle=-90,width=4.2cm]{figb2b.ps}\hspace*{0.1cm}
\includegraphics[angle=-90,width=4.2cm]{figb2c.ps}\hspace*{0.1cm}
\includegraphics[angle=-90,width=4.2cm]{figb2d.ps}\hspace*{0.1cm}
}
\vspace*{0.8cm}
\centerline{
\includegraphics[angle=-90,width=4.2cm]{figb2e.ps}\hspace*{0.1cm}
\includegraphics[angle=-90,width=4.2cm]{figb2f.ps}\hspace*{0.1cm}
\includegraphics[angle=-90,width=4.2cm]{figb2g.ps}\hspace*{0.1cm}
\includegraphics[angle=-90,width=4.2cm]{figb2h.ps}\hspace*{0.1cm}
}
\vspace*{0.8cm}
\centerline{
\includegraphics[angle=-90,width=4.2cm]{figb2i.ps}\hspace*{0.1cm}
\includegraphics[angle=-90,width=4.2cm]{figb2j.ps}\hspace*{0.1cm}
\includegraphics[angle=-90,width=4.2cm]{figb2k.ps}\hspace*{0.1cm}
\includegraphics[angle=-90,width=4.2cm]{figb2l.ps}\hspace*{0.1cm}
}
\vspace*{0.8cm}
\hspace*{-8.8cm}
\centerline{
\includegraphics[angle=-90,width=4.2cm]{figb2m.ps}\hspace*{0.1cm}
\includegraphics[angle=-90,width=4.2cm]{figb2n.ps}\hspace*{0.1cm}
}
\caption{{\bf (cont).}\ Spectra of sources without detected \hi\
absorption. The horizontal lines shown represent the $1\sigma$ standard
deviation from the fitted continuum level. Features due to radio
frequency interference are marked 'RFI'. Velocities are with respect to
the optical redshifts, positive corresponding to infall. The grey vertical
regions indicate the outer 25\% of the total bandwidth.}
\label{fig:nondetb}
\end{figure*}

\begin{figure}[t]
\centering
\centerline{
\includegraphics[angle=-90,width=8.5cm]{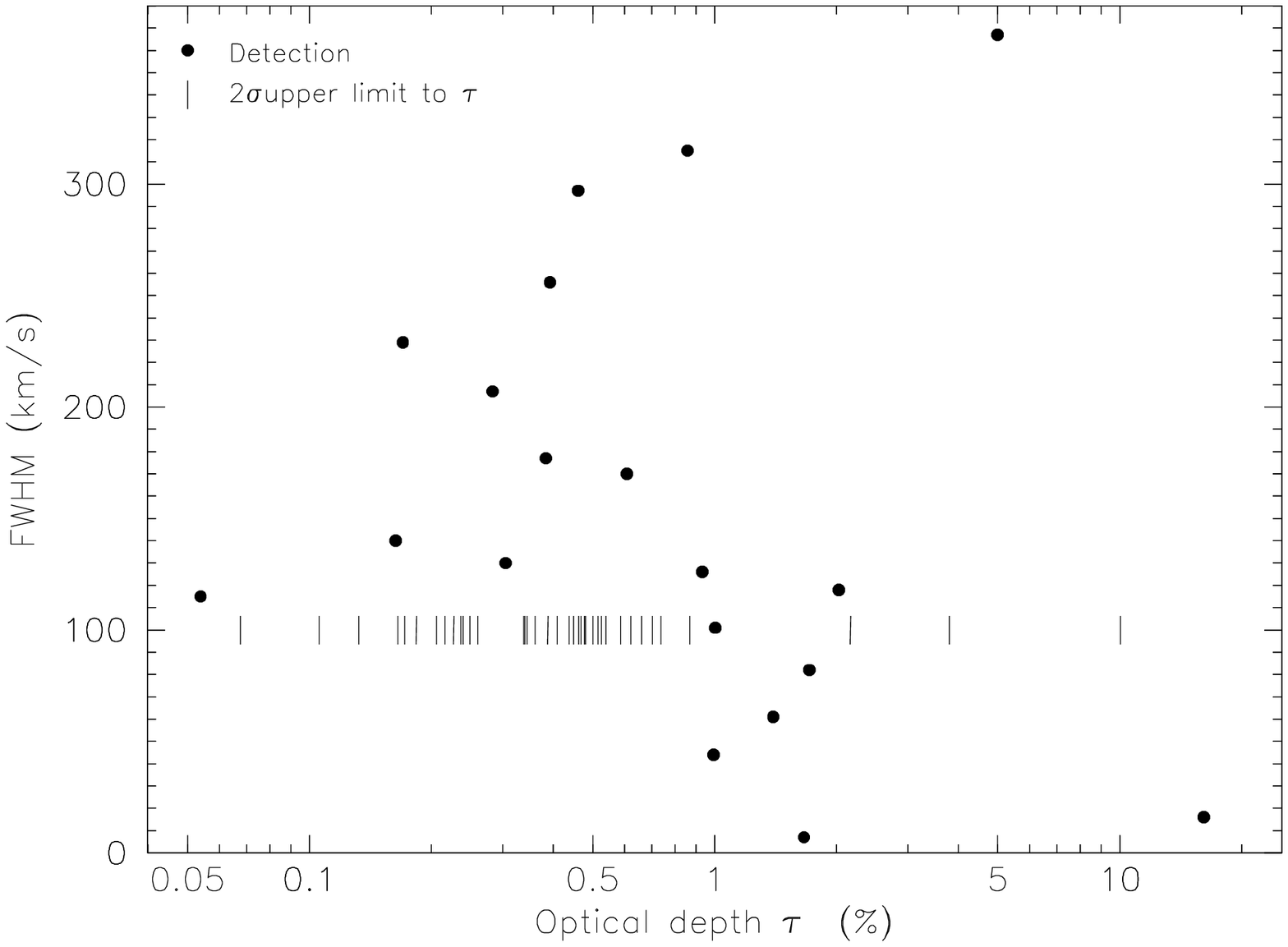}
}
\caption{Distribution of line strengths and widths. The $2\sigma$
non-detection limits are shown as vertical marks at 100\,\kms; any \hi\
absorption in these objects probably has an opacity smaller than (to
the left of) the mark.}
\label{fig:sens}
\end{figure}

\begin{figure}[t]
\centering
\centerline{
\includegraphics[angle=-90,width=8.5cm]{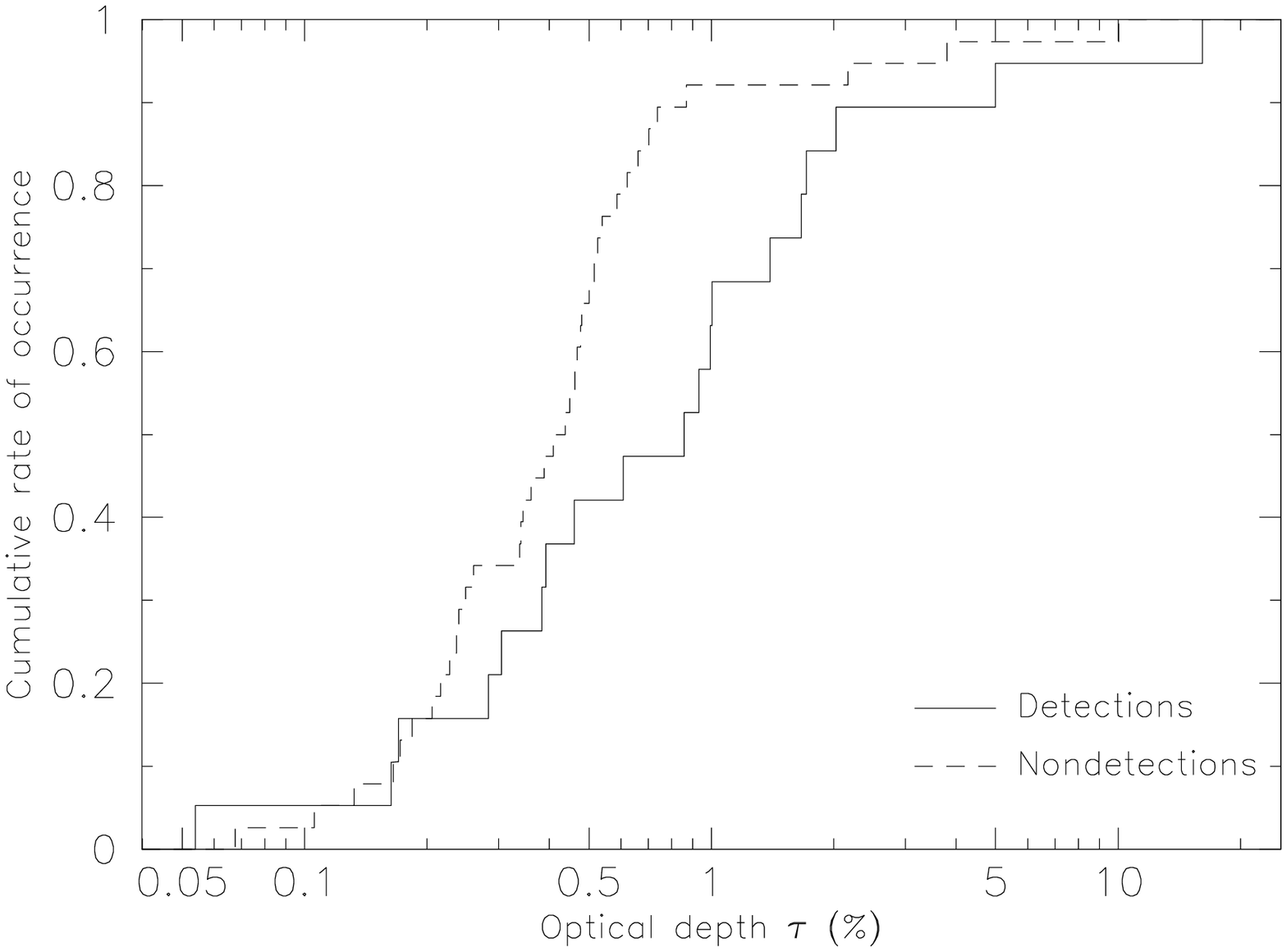}
}
\caption{The cumulative distribution of the opacity of the
detections (solid line) and the non-detections (dashed line), as
discussed in Section~\ref{sec:det}}
\label{fig:cum}
\end{figure}

\begin{figure}[tb]
\centering
\centerline{
\includegraphics[angle=-90,width=8.5cm]{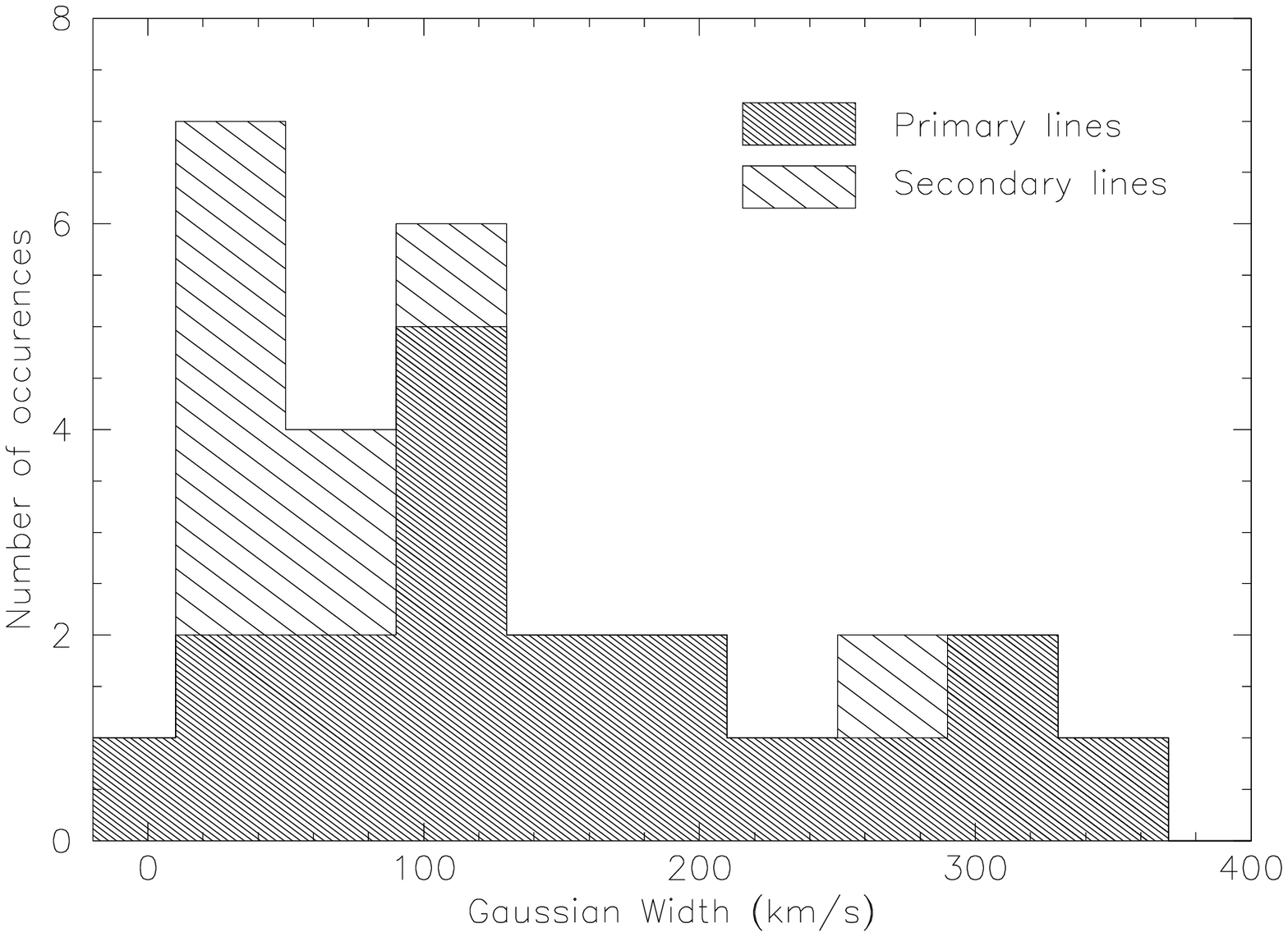}
}
\caption{Number distribution of line widths, in 40\,\kms\ wide bins, as
discussed in Section~\ref{subsec:prop-linew}. The main lines (largest
optical depth in each source) and the secondary lines are indicated
separately.}
\label{fig:fwhm}
\end{figure}

\subsection{Inflowing and outflowing gas ? \label{subsec:prop-flow}}

It was pointed out by van Gorkom et al.\ (1989) that in all 4 out of 29
nearby radio galaxies ($z<0.13$) with detected \hi\ absorption, and
also in 4 other nearby systems, the \hi\ velocity is positive, i.e.\
infalling with respect to the systemic velocity of the galaxy. The
radio structure (or, in the case of 3C236, at least the absorbed part)
of those nearby galaxies with \hi\ absorption is very compact, and many
of these objects are now classified as GPS or CSS sources. Thus, they
potentially match well with the types of sources that constitute the
majority of our sample.

\begin{figure}[thb]
\centering
\centerline{
\includegraphics[angle=-90,width=8.5cm]{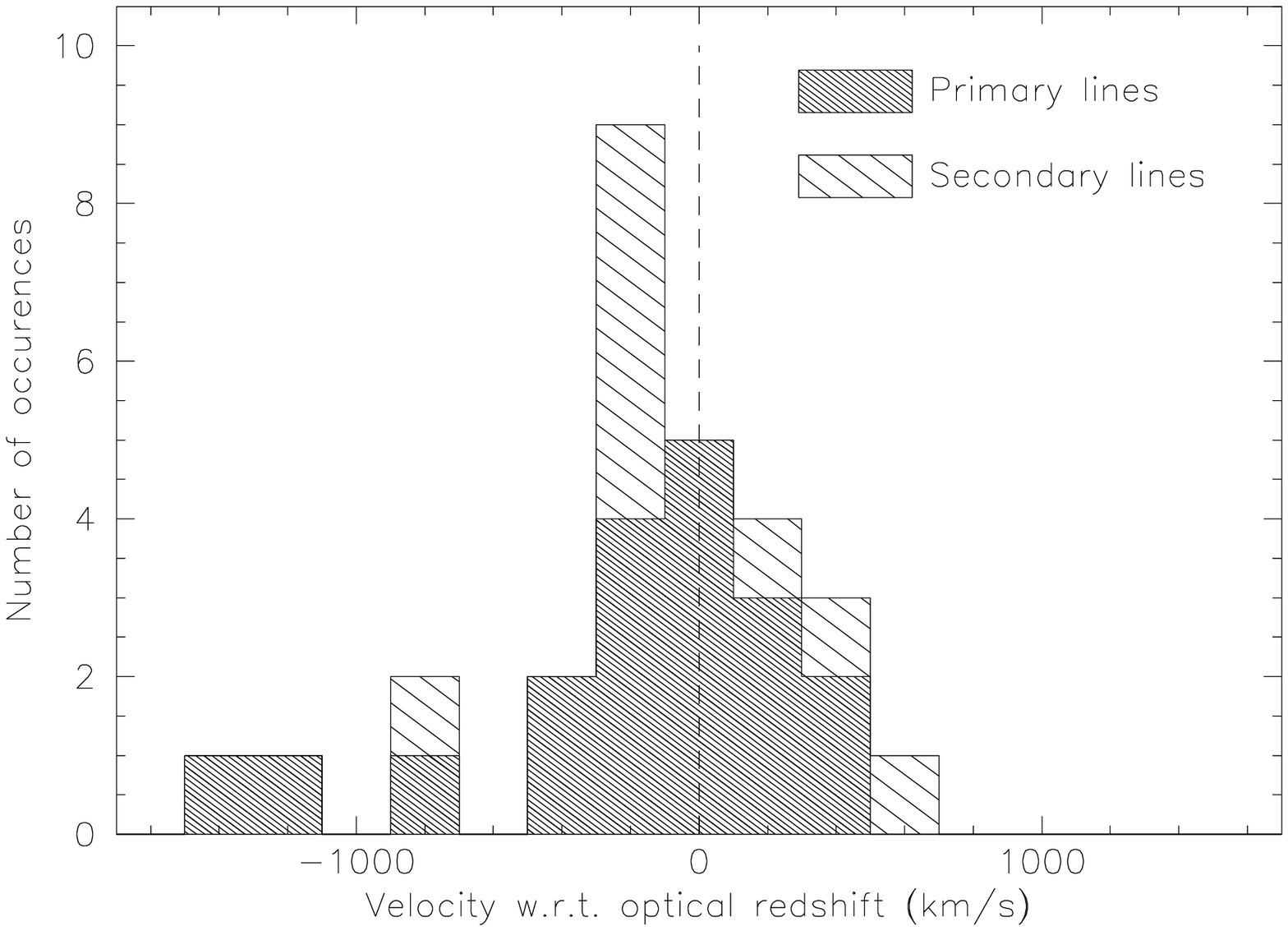}
}
\caption{\hi\ line velocities compared to the optical redshifts, in bins
of width 200\,\kms. The main lines (largest optical depth in each
source) and the secondary lines are indicated separately; their
distributions are discussed in Section~\ref{subsec:prop-flow}.}
\label{fig:vcen}
\end{figure}

As Figure~\ref{fig:vcen} shows, we now find \hi\ lines at both
positive and negative velocities with respect to the optical
redshift. Also, while typical velocities in the lower redshift sources
of van Gorkom (1989) were in the range 0 to $+200$\,\kms, with only 1
value around $+400$\,\kms, 10 of our 19 main (highest opacity) lines
have a velocity more than 200\,\kms\ different from the optical
redshift; the main line velocities range between $v=-1420$\,\kms\ and
$v=+318$\,\kms. While, as already discussed in Section~\ref{sec:det},
inaccuracies in the optical redshift may play a role in causing some
of the scatter, we believe the larger \hi\ velocities are likely to be
significant. We see an intriguing hint that the velocity distribution
may be skewed towards outflow rather than infall. This effect can be
stated more dramatically than it appears through the binning of
Fig.~\ref{fig:vcen}: 13 of the main line velocities are negative as
compared to 6 positive; of these, 7 have $v<-200$\,\kms\ as compared
to 3 having $v>200$\,\kms, and of these, 3 have $v<-600$\,\kms\ as
compared to none at $v>600$\,\kms. The mean centroid velocities are
$v=-223$\,\kms\ for the main lines and $v=-58$\,\kms\ for the
secondary lines, respectively.  However, Student's $t$ tests at a
confidence level of 95\% show that the mean velocities of the main
lines and the secondary lines indeed do not differ significantly from
zero, and also not from each other, possibly only because the sample
size is too small.

Thus, we find the opposite of the results obtained for nearby radio
sources by van Gorkom et al.\ (\cite{vangorkom89}) in both the sign and
the magnitude of the velocities. Speculatively, we think that in these
higher luminosity sources, interaction between the radio jets and the
surrounding inner galactic medium could lead to significant motions of
the gas, such as found in 3C236 by Conway \&\ Schilizzi
(\cite{conway00}). Other evidence for such interactions is often seen
as the so-called radio-optical alignment effect. The kinematics of the
aligned gas seen in CSS sources shows velocity offsets of a few hundred
\kms\ (both positive and negative) with respect to the nuclear velocity
(e.g., O'Dea et al.\ \cite{odea02}). Alternatively, in some cases we
may be looking at the line-of-sight rotational motion of a neutral
component of broad-line gas, seen in front of a parsec-scale radio
source, as has been discussed for lower redshift galaxies for example
by Morganti et al.\ (\cite{morganti01}).

\section{Summary \label{sec:sum}}

\hi\ absorption searches have been carried out with the WSRT for a
sample of mostly compact radio sources with redshifts out to nearly
$z=1$, giving observing frequencies which fall in the UHF-high band
(725--1200~MHz). Out of 57 sources for which adequate spectra could be
obtained, 19 are found to have associated \hi\ in absorption, with
opacities ranging from 16\%\ down to 0.2\%; below 1\% opacity, our
statistics may well be incomplete. The typical line width is $\sim
150$\,\kms, but the lines can also be several times narrower or wider.
Column densities ranging from $1\times 10^{19}$ to $3.3\times 10^{21}$
are obtained by assuming uniform coverage of the radio source and a spin
temperature $T_{\rm sp}=100$\,K. In contrast to nearby radio sources, we
find not only positive but even more negative \hi\ velocities, up to
more than 1000\,\kms (as compared to the optical redshifts). Perhaps,
inflow and outflow can both occur, or there is interaction with the
radio source, or we are seeing rotating broad-line gas.

\begin{acknowledgements}
  The WSRT is operated by ASTRON (The Netherlands Foundation for
  Research in Astronomy) with support from the Netherlands Foundation
  for Scientific Research (NWO). We are very grateful for the dedication
  of the ASTRON staff who worked on the WSRT upgrade, as well as those
  who assisted in performing the survey observations. This research has
  made use of the NASA/IPAC Extragalactic Database (NED), which is
  operated by the Jet Propulsion Laboratory, California Institute of
  Technology, under contract with the National Aeronautics and Space
  Administration in the United States of America.
\end{acknowledgements}

\end{document}